\documentclass[a4paper]{article}

\usepackage{setspace}
\usepackage{url}
\usepackage{hyperref}

\usepackage{bm}
\usepackage{bbm} 
\usepackage{algorithm}
\usepackage{algorithmic}
\usepackage{cancel}

\RequirePackage{authblk}
\RequirePackage{fancyhdr}  
\RequirePackage{lastpage}  
\RequirePackage{times}      
\RequirePackage{mathptmx}   
\RequirePackage{ifpdf}

\RequirePackage{amsmath,amsfonts,amssymb}
\def\xabstract{abstract}
\long\def\abstract#1\end#2{\def\two{#2}\ifx\two\xabstract 
	\long\gdef\theabstract{\ignorespaces#1}
	\def\go{\end{abstract}}\else
\typeout{^^J^^J PLEASE DO NOT USE ANY \string\begin\space \string\end^^J
	COMMANDS WITHIN ABSTRACT^^J^^J}#1\end{#2}
	\gdef\theabstract{\vskip12pt BADLY FORMED ABSTRACT: PLEASE DO
NOT USE {\tt\string\begin...\string\end} COMMANDS WITHIN
THE ABSTRACT\vskip12pt}\let\go\relax\fi
\go}

\usepackage[utf8]{inputenc}
\usepackage[backend=biber,citestyle=nature,bibstyle= chem-biochem, doi=true, maxbibnames=4, minbibnames=1, terseinits=true, url=true]{biblatex}

		\renewbibmacro*{name:last-first}[4]{%
			\ifuseprefix
			{\usebibmacro{name:delim}{#3\isdot}\space#1\isdot} 
			{#1\isdot}
			\ifblank{#4}
			{}
			{\space#4\isdot}
		}

		\DeclareFieldFormat{titlecase}{\MakeSentenceCase*{#1}}  
		\DeclareFieldFormat{doi}{doi\addcolon\href{https://doi.org/#1}{\nolinkurl{#1}}} 
		
		\usepackage{xpatch}
		\xpretobibmacro{author}{\mkbibbold\bgroup}{}{}
		\xapptobibmacro{author}{\egroup}{}{}
		\xpretobibmacro{bbx:editor}{\mkbibbold\bgroup}{}{}
		\xapptobibmacro{bbx:editor}{\egroup}{}{}

		\renewcommand*{\andothersdelim}{\addcomma\addspace} 
		
		\renewbibmacro*{name:andothers}{
			\ifboolexpr{
				test {\ifnumequal{\value{listcount}}{\value{liststop}}} 
				and
				not test \ifmorenames 
			}
			{\adddot} 
			{}%
			\ifboolexpr{
				test {\ifnumequal{\value{listcount}}{\value{liststop}}}
				and
				test \ifmorenames 
			}
			{\ifnumgreater{\value{liststop}}{1}
				{\finalandcomma} 
				{}%
				\andothersdelim\bibstring[\emph]{andothers}}
			{}%
		}

		\usepackage{graphicx}
		\usepackage{float} 
		\usepackage{longtable}
		\usepackage{multirow}
		\usepackage{caption}
		\usepackage{subcaption}
		\usepackage{mathtools}
		\usepackage{indentfirst}

		\usepackage[running]{lineno}
		\usepackage{csquotes}

		\newcommand*\patchAmsMathEnvironmentForLineno[1]{
			\expandafter\let\csname old#1\expandafter\endcsname\csname #1\endcsname
			\expandafter\let\csname oldend#1\expandafter\endcsname\csname end#1\endcsname
			\renewenvironment{#1}
			{\linenomath\csname old#1\endcsname}
			{\csname oldend#1\endcsname\endlinenomath}}
		\newcommand*\patchBothAmsMathEnvironmentsForLineno[1]{
			\patchAmsMathEnvironmentForLineno{#1}
			\patchAmsMathEnvironmentForLineno{#1*}}
		\AtBeginDocument{
			\patchBothAmsMathEnvironmentsForLineno{equation}
			\patchBothAmsMathEnvironmentsForLineno{align}
			\patchBothAmsMathEnvironmentsForLineno{flalign}
			\patchBothAmsMathEnvironmentsForLineno{alignat}
			\patchBothAmsMathEnvironmentsForLineno{gather}
			\patchBothAmsMathEnvironmentsForLineno{multline}
		}

		\title{Modelling immunity in agent-based models\\[0.5ex]
		}
		
		\begin{abstract}
			
		\end{abstract}

\author[1,2,4,5,*]{Gray Manicom}
\author[1,3]{Emily Harvey}
\author[2,6]{David Wu}
\author[2,7,8]{Joshua Looker}
\author[10]{Oliver Maclaren}
\author[1,2,9]{Dion R.J. O'Neale}

\affil[1]{Te P\={u}naha Matatini, New Zealand}
\affil[2]{Department of Physics, University of Auckland, New Zealand}
\affil[3]{M.E. Research, Takapuna, Auckland, New Zealand}
\affil[4]{Policy Innovation Lab, Stellenbosch University, South Africa}
\affil[5]{National Institute for Theoretical and Computational Sciences (NITheCS) South Africa, South Africa}
\affil[6]{Department of Econometrics and Business Statistics, Monash University, Australia}
\affil[7]{EPSRC and MRC Centre for Doctoral Training in Mathematics for Real-World Systems, University of Warwick, United Kingdom}
\affil[8]{Zeeman Institute for Systems Biology and Infectious Disease Epidemiology Research, University of Warwick, United Kingdom}
\affil[9]{Nicholson Consulting, Wellington, New Zealand}
\affil[10]{Department of Engineering Science and Biomedical Engineering, University of Auckland, New Zealand}
\affil[*]{Corresponding author e-mail addresses: graym@sun.ac.za}

\begin{document}

\maketitle
\raggedbottom
\thispagestyle{fancy}
\setcounter{page}{1}
\fancyfoot[R]{Page \textbf{\thepage}}
\pagestyle{fancy}
\fancyhf{}

\newpage

\section{Summary}
Vaccination policies play a central role in public health interventions and models are often used to assess the effectiveness of these policies. Many vaccines are leaky, in which case the observed vaccine effectiveness depends on the force of infection. Within models, the immunity parameters required for agent-based models to achieve observed vaccine effectiveness values are further influenced by model features such as its transmission algorithm, contact network structure, and approach to simulating vaccination. We present a method for determining parameters in agent-based models such that a set of target immunity values are achieved. We construct a dataset of desired population-level immunity values against various disease outcomes considering both vaccination and prior infection from COVID-19. This dataset incorporates immunological data, data collection methodologies, immunity models, and biological insights. We then describe how we choose minimal parameters for continuous waning immunity curves that result in those target values being realized in simulations. We use simulations of the household secondary attack rates to establish a relationship between the protection per infection attempt and overall immunity, thus accounting for the dependence of protection from acquisition on model features and the force of infection.
\newpage

\fancyfoot[R]{Page \textbf{\thepage}}
\renewcommand{\headrulewidth}{2pt}
\renewcommand{\footrulewidth}{1pt}

\section{Introduction}
Infectious diseases continue to pose significant challenges to global health, necessitating robust models to inform decision-making and evaluate intervention strategies. Central to these models is the accurate representation of immunity, which is dynamic due to factors such as waning over time and the emergence of new pathogen variants. Vaccination campaigns aim to develop herd immunity and mitigate the spread of diseases, making it critical to obtain reliable estimates of immunity from both vaccination and previous infections for incorporation into mathematical models.

The force of infection is the rate at which susceptible individuals become infected in a given time window. Vaccine effectiveness (VE) measures how much that risk is reduced among vaccinated individuals in observational studies. However, many vaccines are leaky, and measures of vaccine effectiveness for leaky vaccines vary in different settings \cite{langwigLimitedAvailableEvidence2019, edlefsenLeakyVaccinesProtect2014}. In particular, vaccine effectiveness measurements tend to decrease as the force of infection increases, with an explanation being that there may be more infection attempts on vaccinated individuals and thus a higher chance of being infected, even when the vaccine blocks certain infection attempts \cite{kaslowForceInfectionDeterminant2021}. Thus, biological factors and non-pharmaceutical interventions, such as wearing masks or social distancing, that impact the force of infection may result in larger measurements of vaccine effectiveness. 

Likewise, in agent-based models of infectious diseases, infectious individuals may make multiple attempts to infect vaccinated susceptible individuals, and the number of these attempts per unit time is determined by the model's transmission mechanism, force of infection, contact structure, implementation of non-pharmeceutical interventions, and other modelling assumptions. To state it simply, the model's representation of VE depends on the model itself. Careful consideration must therefore be made in choosing the immunty parameters that will result in the desired VE being observed in model simulations. In particular, this means determining the relationship between the protection that immunity provides against each infection attempt and the overall immunity of the simulated population. Additionally, the immunity that leaky vaccines provide tends to wane over time, and so the protection against infection attempts and subsequent disease outcomes should also wane. Various approaches have been used to model leaky immunity for different models based on experimental data. In compartment models, this has been done by having a rate at which individuals move from a recovered state to a susceptible state \cite{aruffoCommunityStructuredModel2022, inayaturohmatMathematicalModelCOVID192021} or by having different compartments for different levels of immunity over time \cite{vattiatoAssessmentPotentialImpact2022}. In agent-based models, this has been done by changing the probability of individual agents entering different disease states, typically introducing new states for different levels of immunity \cite{truszkowskaPredictingEffectsWaning2022,chenRoleHeterogeneityNational2024,hinchOpenABMCovid19agentbasedModelNonpharmaceutical2021,kerrCovasimAgentbasedModel2021,roubenoffHowWillCOVID192023,barmpounakisEvaluatingEffectsSeconddose2022,pillaiAgentbasedModelingCOVID192024}. 

The significance of this research is the demonstration of how vaccine effectiveness (VE) or immunity from previous infections, observed in epidemiological studies, can be accurately translated into model parameters, accounting for the complexities of how the immunity provided by leaky vaccines depends, in the real world and in simulations, on the force of infection and modeling assumptions. This makes the model more accurate, relevant and flexible for real-world applications. This paper describes a method for incorporating immunity against disease acquisition, symptoms, hospitalization, critical care, and death into an agent-based model, using COVID-19 as a case study. Immunity results from vaccination with one to four doses of the Pfizer vaccine with or without previous infection, giving nine classes of immunity for each disease outcome which wane over time. Unlike many agent-based models, which introduce new states where individuals may have fixed immunity levels for long periods, our approach represents immunity as a continuous time-dependent multiplier on the rate of infection. This continuous representation is advantageous for modelling scenarios with variable intervals between vaccinations and reinfections and accommodates different waning rates for various disease outcomes. 

To account for how vaccine effectiveness depends on modelling assumptions and the force of infection, we set up a series of numerical simulations to act as case-control studies, use these simulations to calculate the relationship between model parameters and vaccine effectiveness, and then determine the parameters that are needed to match target vaccine effectiveness values from the literature. To measure vaccine effectiveness in simulations, we run simplified numerical simulations of spread in disconnected households. We then use the household secondary attack rate (HSAR), which is the probability that a susceptible individual in a household gets infected when at least one other individual in that household gets infected \cite{hiltonComputationalFrameworkModelling2022}, to determine the parameter needed for `protection against acquisition per infection attempt' to produce overall population-level vaccine effectiveness values for different infection pressures. This gives the model accuracy and flexibility when applied to real-world modelling scenarios, since different policy interventions and biological adaptations may result in the force of infection changing, which, in turn, changes the relationship between model parameters and the population-level immunity observed in simulations.

In Subsection \ref{subsec:model_description} we introduce one of the agent-based models used as part of the COVID-19 response in New Zealand. In Section \ref{sec:methods} we describe the methods we use to acquire parameters for our model that produce realistic protection values. In Subsections \ref{subsec:design} and \ref{subsec:notation} we overview the method and introduce notation. In Subsection \ref{subsec:target_dataset} we construct a target dataset of immunity values based on a model fit to observed data, making adjustments based on biological insights, observed data, and modelling considerations. In Subsection \ref{subsec:v_to_V_methods} we provide the method for determining the parameters of logistic transition probability curves that can be incorporated into our model to achieve the target immunity values in simulations. Theoretical and numerical analysis is used to determine the how to fit model parameters to the target dataset with special attention paid to the relationship between the protection from acquisition per infection attempt and the cumulative protection. In Section \ref{sec:results} we show the results of this method, first by showing the target immunity values in Subsection \ref{subsec:target_dataset_results} and then by showing the immunity values that are achieved in model simulations in Subsection \ref{subsec:parameterisation_results}. Finally, in Section \ref{sec:discussion} we discuss the results.

\subsection{The Network Contagion Model of New Zealand} \label{subsec:model_description}

One part of New Zealand's COVID response was the use of an agent-based network contagion model (NCM) to predict the impact of certain policies \cite{mccawROLEMATHEMATICALSCIENCES2022}. The NCM simulates the spread of the disease on a bipartite contact network built using a synthetic population representative of New Zealand; Each individual is represented by a node in the network and there are group nodes that represent different interaction contexts. Individuals and groups in the network have certain attributes; for example, individuals will have an age, ethnicity and vaccination date attribute that provides us with knowledge of both how many vaccinations they have received and when they received them \cite{NzcoviddataVaccinedataMain}. Infections may occur when an infectious individual is connected to the same group node as a susceptible one. 

\begin{figure}[ht!]
	\centering
        \includegraphics[width=0.6\linewidth]{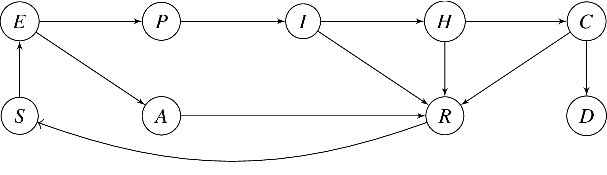}
	\caption{Possible transitions between different disease states for the network contagion model. An individual in the $A$, $P$ and $I$ states may infect someone in the $S$ state if they are contacts. Recovered individuals become susceptible again, but will have some immunity from previous infection.} 
	\label{fig:stateChanges}
\end{figure}

The probability of infection primarily depends on disease characteristics, the contact structure of the network, policies that affect the rate of spread in different interaction contexts, the attributes of the individuals, and their immunity from vaccination or previous infection. Infected individuals may then transition to more advanced disease states with probability dependent on the individual's attributes. The disease states are Susceptible ($S$), Exposed ($E$), Asymptomatic ($A$), Presymptomatic ($P$), Infectious ($I$), Hospitalized ($H$), Critical Care ($C$), Dead ($D$) and Recovered ($R$). We call the conditional probability of transitioning from state $X$ to state $Y$ a transition probability and the associated transition diagram is shown in Figure \ref{fig:stateChanges}. Control states represent the effect of non-pharmaceutical interventions, but since they are independent of immunity, we ignore them in this work.

A key modeling assumption in agent-based disease models is whether the transmission is density-dependent or frequency-dependent. In density-dependent models (often called “mass-action” models), the contact rate between individuals (and hence the transmission rate) scales with population size. As the population grows larger, each individual has more contacts overall, which increases the total number of infection attempts. In frequency-dependent models, by contrast, as the population size increases, each infectious individual still makes the same number of infection attempts, so the key factor is what fraction of the population is infected rather than the total number of contacts. In our model, we assume a frequency-dependent spread. Other model choices are that individuals in $A$ and $P$ are infectious, but less so than individuals in $I$; and that the time spent in each state has an exponential distribution except for the time spent in $E$, which has a gamma distribution. To efficiently simulate the non-Markovian dynamics that arise from these choices, including the time-dependent effect of waning immunity presented in this paper, we use a thinning-based method, similar to that presented in \cite{grossmannEfficientSimulationNonMarkovian2020}. 

By introducing dependencies between the acquisition and development of the disease with individual and group characteristics, this model can capture significant heterogeneous features that compartment models cannot \cite{siegenfeldModelingComplexSystems2022}. However, this comes at the cost of additional complexity within the model and the need for greater computational power. Our approach of parsimoniously parametrizing immunity from vaccines without introducing new compartments or disease states reduces the complexity and computational load while maintaining the heterogeneous characteristics of the NCM. Code repositories and documentation for constructing the synthetic household contact network \cite{emilyharveyHouseholdSyntheticNetwork2024} and the network contagion model are available at \cite{cobin_documentation}.

\section{Methods}\label{sec:methods}


\subsection{Overview} \label{subsec:design}

In this section, we describe a method for determining the effect of immunity on the probability of susceptible individuals acquiring the disease, as well as transitioning from the exposed state to symptomatic, symptomatic to hospitalized, hospitalized to critical care, and critical care to death. Our approach is to fit three-parameter logistic curves to a target dataset for each of the nine different classes of immunity and five different transition probabilities. We describe the approach, introduce notation, discuss how we construct a target dataset of immunity values, and then describe the method for calculating the per-attempt protection from disease acquisition that immunity provides. 

Particular attention is given to determining the role of vaccination in protecting susceptible individuals from acquiring the disease. Estimates of vaccine effectiveness in both the real world and in the results of model simulations will depend on a wide range of factors. In particular, since leaky vaccines allow for multiple infection attempts on vaccinated susceptible individuals, the simulation factors that determine the number of infection attempts on susceptible individuals need to be taken into consideration. These include the underlying force of infection, the contact structure, and the transmission mechanism. Therefore, the values of vaccine effectiveness used in our model will not be equal to the values observed in the real world, and appropriate model parameters need to be chosen that realize the target observed values. 

Attempting to theoretically calculate the relationship between the model parameters and the vaccine effectiveness observed in model simulations quickly becomes intractable due to the exponentially increasing number of transmission pathways and the large number of transmission settings. If the pressure of infection is low then the probability of seeing long transmission pathways becomes small and simplifying assumptions can be made, but in our case these assumptions do not apply \cite{sharkerEstimatingInterpretingSecondary2021}. The method we use to solve this issue is to use numerical simulations to determine the relationship between the input immunity parameters and the resulting population immunity observed in the model. 

In our first experiment we use simulations of a simple scenario where each infectious individual, who have a fixed infectivity, has only one other contact, who is susceptible but with some `protection against infection per infection attempt'. We run this scenario $140675$ times to determine how the susceptible individual's `protection per infection attempt' affects the overall probability of them getting infected. This simple scenario allows us to calculate the relationship theoretically and verify the simulations. We repeat this over a range of different values for the force of infection. Individuals in this simulation are sampled from a synthetic population and have different attributes such as age and ethnicity. These attributes affect the infectiousness of individuals through differences in disease severity, which we need to account for in our theoretical calculations. However, we set parameters such that the susceptibility to infection does not vary between susceptible individuals.

In our second experiment we run simulations where an infected individual is in an isolated household of susceptible contacts and the household size is sampled from the distribution of household sizes of New Zealand. These results are used to calculate the HSAR and determine the relationship between immunity per infection attempt and overall observed vaccine effectiveness. We use the household as a transmission unit for several reasons: Firstly, when many of the vaccine effectiveness studies were done worldwide, there were intensive testing, contact tracing, isolation, and quarantine interventions in place and thus most spread was in households. Secondly, households are the most common transmission unit used to determine vaccine effectiveness since it is usually easy to identify the original infectors and their susceptible contacts \cite{halloranDesignAnalysisVaccine2010}, and some of the studies used for constructing our target values used the HSAR \cite{lyngseSARSCoV2OmicronVOC2021,singanayagamCommunityTransmissionViral2022,harrisEffectVaccinationHousehold2021a}. Thirdly, by using the HSAR we could avoid having to make assumptions about the other interaction contexts and about the role of non-pharmaceutical interventions. This allowed us to run experiments seeding infection in a random 5\% of households simultaneously on an existing synthetic household-only network \cite{emilyharveyHouseholdSyntheticNetwork2024}. By running this simulation five times (seeding into the same $65489$ households) we end up with $327445$ independent simulations, which we repeat over a range of different infection pressure and protection against infection parameter values. 

For ease in setting up the simulations, and to compare the results from the two sets of experiments, the disconnected pairs of contacts in the first experiment are simulated using the same synthetic household network as in the second experiment, but with infections only seeded in a selection of disconnected two-person households. We measure the HSAR as the fraction of individuals who were susceptible at the experiment’s outset and became infected by its conclusion. 

 \subsection{Notation} \label{subsec:notation} 
 Let $\mathbb{S}$ be the set of all disease states. Let $t$ be the period between the current time and the time of the most recent vaccination or recovery and $i=1,...,9$ represent one of the nine combinations of vaccine doses with or without previous infection. Let $X \in \mathbb{S}$ be an incoming state of $Y \in \mathbb{S}$ and write the corresponding transition probability as $P(Y \; | \; X)(t)$, the reducing factor on the transition probability for immunity from combination $i$ as $f^{(i)}_{XY}(t)\in [0,1]$, and the transition probability in the case where there is no immunity by $\eta_{XY}$. Then 
 \begin{align*}
    P\big(Y \mid X \big)(t)
    &= f^{(i)}_{XY}(t) \eta_{XY}.
\end{align*}
Let $V^{(i)}_{XY}(t)$ denote the overall protection against transitioning from $X$ into $Y$ due to combination $i$ (this is the vaccine effectiveness measured in population studies in cases where immunity is due to vaccination). In our model, the value $\eta_{XY}$ depends on the disease infectivity and individual attributes. Note that $V^{(i)}_Y(t)$ typically decreases (as immunity wanes) while $f^{(i)}_{XY}(t)$ typically increases (as the probability of becoming more ill increases) for each immunity $i$. 

Since we are interested in the effect of immunity on the probability that infected individuals enter more severe disease states, we focus our attention on the disease states $\mathbb{T}=\{E,P,H,C,D\}$ and simplify the notation based on the following observations: Firstly, for each $Y\in \mathbb{S}\setminus \mathbb{T}$ we assume that $V_{XY}=0$ and $f_{XY}=1$. Thus we only need to find immunity values for $Y\in \mathbb{T}$. Secondly, all individuals who enter $P$ will transition to $I$, regardless of immunity, and therefore we do not include $I$ in $\mathbb{T}$ and we treat $P$ as the incoming state of $H$, where appropriate, for the sake of calculating immunity. Thirdly, since each $Y\in \mathbb{T}$ has only one incoming state $X$, we omit the incoming state. Finally, since the methodology is applied uniformly to all immunities, we focus on a single immunity and omit the superscript $(i)$. Thus we write
\begin{align*}
P\big(Y \mid X \big)(t) &= f_Y(t) \eta_Y
\end{align*}
for each $Y\in \mathbb{T}$ where
\begin{equation}\label{eq:transition_prob}
    f_{Y}(t)=\frac{1-V_Y}{1-V_X}
\end{equation}
for each $Y\in \mathbb{T} \setminus \{E\}$. We discuss how we determine $V_Y$ in the next section and then show how we calculate $f_{Y}$ given $V_{Y}$ in Section \ref{subsec:v_to_V_methods}.

\subsection{Constructing a target dataset} \label{subsec:target_dataset} 

This section describes how we construct our target dataset of immunity values from vaccines and previous infections against the Omicron variant of COVID-19. Estimates of vaccine effectiveness for leaky vaccines are obtained with either control or cohort studies, and the estimates may vary with context, methodological choices, and disease characteristics such as the force of infection \cite{smithAssessmentProtectiveEfficacy1984,halloranDesignAnalysisVaccine2010}. 

Our starting point is a model of Khoury et. al. that establishes a relationship between in vitro neutralizing antibody levels and immune protection against symptomatic infection, aiming to identify correlates of protection and predict vaccine efficacy \cite{khouryNeutralizingAntibodyLevels2021}. Golding fit this model to provide curves of waning immunity to acquisition of different variants of COVID-19 \cite{goldingNeuts2efficacySourceCode2022}. We used Golding's dataset released on the 5\textsuperscript{th} April, 2022, for immunity against the Omicron BA.1 and BA.2 strains. Where applicable, we adjust the model values for two reasons: first, if there is more up-to-date data concerning vaccine effectiveness \cite{ukhealthsecurityagencyCOVID19VaccineSurveillance2022b,chemaitellyDurationImmuneProtection2022,andrewsCovid19VaccineEffectiveness2022,liuVaccinesElicitHighly2022,altarawnehEffectsPreviousInfection2022,tartofDurabilityBNT162b2Vaccine2022,baumHighVaccineEffectiveness2022,mcmenaminVaccineEffectivenessOne2022,hansenRiskReinfectionVaccine2023,ukhealthsecurityagencyCOVID19VaccineSurveillance2022} and secondly, if there are other biological insights which apply to waning immunity that were not included in his model, but which we think are significant \cite{khouryNeutralizingAntibodyLevels2021,setteImmunologicalMemorySARSCoV22022,szanyiLogoddsSystemWaning2022,priddyImmunogenicityBNT162b2COVID192022}. We summarize these results below:
\begin{itemize}
    \item[1)] We begin by assuming that the protection against acquisition and developing symptoms is primarily determined by neutralizing antibodies produced in response to vaccination or previous infection. That is, Golding's model of waning immunity is a good baseline for determining immunity over time, even if it does not consider all factors contributing to immunity \cite{markingHighRateBA12022}.
    \item[2)] We assume that immunity provides higher protection against being admitted into critical care than being hospitalized, that is, $V_C(t)>V_H(t)$ and $f_{C}(t)<1$. This is based on observations in the UK \cite{ukhealthsecurityagencyCOVID19VaccineSurveillance2022} 
     (where `critical care' refers to someone admitted into secondary care for at least $2$ days due to a respiratory illness and requires oxygen, ventilation or ICU) and Finland \cite{baumHighVaccineEffectiveness2022} (where `critical care' refers to ICU admission) which show different levels of protection from vaccination between hospitalization and critical care.
    \item[3)] Once infected, the immune response of immune cells other than neutralizing antibodies, such as plasmablast, memory B and T cells, provide further protection \cite{scottCovid19VaccinationEvidence2021,setteImmunologicalMemorySARSCoV22022}. The protection provided by these cells is robust and long-lasting so these immune responses provide individuals who have been vaccinated or previously infected long-term protection against severe disease. Thus we assume that the protection against hospitalization, critical care, and death given infection will wane to a certain point, but will wane no further. In our model, this corresponds to us adjusting the protection against severe disease provided by Golding so that $f_H(t)$, $f_C(t)$, and $f_D(t)$ stop increasing and are constant after $5$ months. Note that $V_H(t)$, $V_C(t)$, and $V_D(t)$ will still decrease, but at a slower rate determined by $V_P(t)$. 
    \item[4)]  An assumption in our model is that all individuals who die must have been in critical care first. Thus, since $V_C(t)$ is already very large, we set
     $$V_D(t) = V_C(t) $$ 
     and assume
     $$ V_C(t)\geq V_H(t) \geq V_P(t) \geq V_E(t)$$
     for all $t\geq 0$. 
    \item[5)] We assume that the protection from onward transmission due to vaccination and/or previous infection is negligible. That is, those already infected despite some form of immunity are not less infectious than those who had no immunity to begin with. 
    \item[6)] Where applicable we adjust Golding's curves to better match more up-to-date data. For example, we increased the protection from hospitalization from the second dose of the Pfizer vaccine from what the model had estimated.
\end{itemize}

\subsection{Protection against acquisition per infection attempt} \label{subsec:v_to_V_methods}

We now describe our method of calculating functions $f_Y$ so that our model will realize our target values $V_Y$ for each $Y\in \mathbb{T}$. We choose $f$ to be a three-parameter logistic function
\begin{equation} \label{eq:logistic}
f_{Y}(t;L_Y,k_Y,T_Y)=\frac{L_Y}{1+\text{exp}\big(-k_Y(t-T_Y)\big)}
\end{equation}
 since this introduces only a few parameters and has been shown to fit waning vaccine efficacy profiles \cite{khouryNeutralizingAntibodyLevels2021,halloranDesignAnalysisVaccine2010}. We fit these curves directly to the target immunity dataset. However, the total immunity against severe disease states depends on all the conditional transition probabilities that lead up to that disease state. Thus we fit the transition probabilities to a target dataset recursively as follows:
\begin{equation}\label{eq:recursive}
\begin{aligned}
    V_D(t)&=1-f_{D}(t;L_D,k_D,T_D)[1-V_C(t)] \\ 
    V_C(t)&=1-f_{C}(t;L_C,k_C,T_C)[1-V_H(t)] \\
    V_H(t)&=1-f_{H}(t;L_H,k_H,T_H)[1-V_P(t)] \\
    V_P(t)&=1-f_{P}(t;L_P,k_P,T_P)[1-V_E(t)].
\end{aligned}
\end{equation}
Note that $P(I \mid P)=1$, thus we consider the protection from hospitalization being dependent on being in $P$, not $I$. 

We can fit $f_D$, $f_C$, $f_H$ and $f_P$ directly to the target dataset. This is because it takes only one random sample to determine whether or not individuals transition from infected to symptomatic, symptomatic to hospitalized, and so on. If they do not transition to a more severe disease state then they recover. However, when vaccines are leaky, there may be multiple infection attempts between infectious and vaccinated susceptible individuals, and when an infection attempt fails the contact remains susceptible. Therefore, many random samples may be used to determine whether individuals get infected. Accordingly, $f_E(t)$ needs to be calculated per contact. By contrast, the values for $V_E$ that we use (in subsection \ref{subsec:target_dataset}) measure the immunity of populations whose level of interaction depends on a wide range of factors. 

We want to choose parameters for $f_E$ so that the protection against acquisition over all contacts matches the target cumulative vaccine effectiveness $V_E$. The theoretical calculation of this relationship is complex since the number of contacts between infectious and susceptible individuals depends on the transmission algorithm, the network's contact structure, the disease characteristics, and other model properties. Therefore, we rely on model simulations of disconnected groups (in this case households) and calculations of the HSAR to determine the per-infection attempt immunity from $V_E$. We validate these simulation results with theoretical calculations of the simple case where households contain one infectious and one susceptible individual. 

Let 
\begin{equation}\label{eq:v}
v(t)=1-f_{E}(t;L_E,k_E,T_E)
\end{equation}
be the protection from acquisition due to vaccination per infection attempt over time. Let $X$ be an infectious state, that is, $X\in\{A,P,I\}$. Let $\beta_X$ be the rate of infection (number of infection attempts per infector per unit time) for an infector in state $X$ and $\alpha_{XY}$ be the transition rate from state $X$ to state $Y$. In our model, $\beta_X$ depends on the control states (non-pharmeceutical interventions) and disease characteristics. The model's assumption of frequency-dependence means that the infector makes the same number of infection attempts per unit time regardless of the number of contacts. That is, if $N$ is the number of contacts of the infector, then the infection rate per contact is $\beta_X/N$. Let $H(v(t))$ (resp. $H_0$) denote the HSAR when all susceptible individuals in the household are vaccinated with the same immunity (resp. unvaccinated). The aggregated immunity $V_E$ is then given as \cite{halloranDesignAnalysisVaccine2010}
\begin{equation} \label{eq:VE_from_HSAR}
V_E(t)=1-\frac{H(v(t))}{H_0}.
\end{equation} 

We begin by investigating a simple case using the following assumptions. 
\begin{itemize}
    \item[(a)] Each individual has only one contact.
    \item[(b)] There are no reinfections.
    \item[(c)] Each individual has the same protection from vaccination, and it remains constant over time, that is, $v(t)=v$ for all $t\geq 0$.
\end{itemize}
Under assumption $(a)$, the HSAR is the probability that the susceptible contact of a seed case becomes infected before the infectious seed case recovers or is hospitalized. We calculate this probability theoretically. 

Since infection attempts against vaccinated susceptible contacts are blocked with constant probability $(1-v)$, the infection rate from infectious state $X$ is $\beta_X (1-v)$ for all time. Consider an individual in the infectious asymptomatic state $A$ with a single susceptible contact. The probability that they infect their susceptible contact before recovery is 
\begin{equation}\label{eq:P_A}
    P_A(v)=\frac{\beta_A(1-v)}{\beta_A(1-v) + \alpha_{AR}}
\end{equation}
since the time to recovery and infection are competing exponentially distributed processes. Similarly, for individuals in the infectious state $P$, the probability that they infect the susceptible contact before transitioning into $I$ is
\begin{equation}\label{eq:P_P}
    P_P(v)=\frac{\beta_P(1-v)}{\beta_P(1-v) + \alpha_{PI}}.
\end{equation}
From $I$, individuals can be hospitalized or recovered. Although the corresponding rates may be different (that is, $\alpha_{IR}$ may be different to $\alpha_{IH}$), the model does not treat these as competing processes. Instead, individuals in $I$ are assigned to transition to $H$ or $R$ with probabilities $\eta_h$ and $1-\eta_h$ when they enter $I$. Thus the probability that an individual in $I$ attempts to infect someone before they are either hospitalized or recovers is 
\begin{equation}\label{eq:P_I}
    P_I(v)=\eta_h \frac{\beta_I(1-v)}{\beta_I(1-v)+\alpha_{IH}} + (1-\eta_h)\frac{\beta_I(1-v)}{\beta_I(1-v)+\alpha_{IR}}.
\end{equation}
The probability that an exposed individual with no immunity becomes pre-symptomatic is $\eta_P$. Thus the HSAR for the simple case is given by 
\begin{equation}\label{eq:HSAR_v}
H(v)=(1-\eta_P) P_A(v) + \eta_P P_P(v) + \eta_P (1-P_P(v)) P_I(v).
\end{equation}
For unvaccinated contacts we can let $v=0$ in equations (\ref{eq:P_A}), (\ref{eq:P_P}) and (\ref{eq:P_I}) to get
\begin{equation}\label{eq:HSAR_u}
H_0=(1-\eta_P) P_A(0) + \eta_P \bigg[P_P(0) + (1-P_P(0)) P_I(0)\bigg].
\end{equation}
Substituting (\ref{eq:v}), (\ref{eq:HSAR_v}) and (\ref{eq:HSAR_u}) into (\ref{eq:VE_from_HSAR}) gives the relationship between $V_E(t)$ and $f_E(t)$ under our simplifying assumptions. This relationship depends on disease characteristics such as infectivity ($\beta_X$) and disease timeline ($\alpha_{XY}$). 

\section{Results} \label{sec:results}
In this section, we implement the techniques discussed in Section \ref{sec:methods} and show the results. We begin by showing the target immunity curves we acquire following the assumptions outlined in Section \ref{subsec:target_dataset}. We then use this target dataset to fit our parameters using equations (\ref{eq:recursive}) and simulations of the HSAR as described in Section \ref{subsec:v_to_V_methods}.

\subsection{Constructing a dataset} \label{subsec:target_dataset_results}
\begin{figure}[th!]
    \centering
        \includegraphics[width=0.95\linewidth]{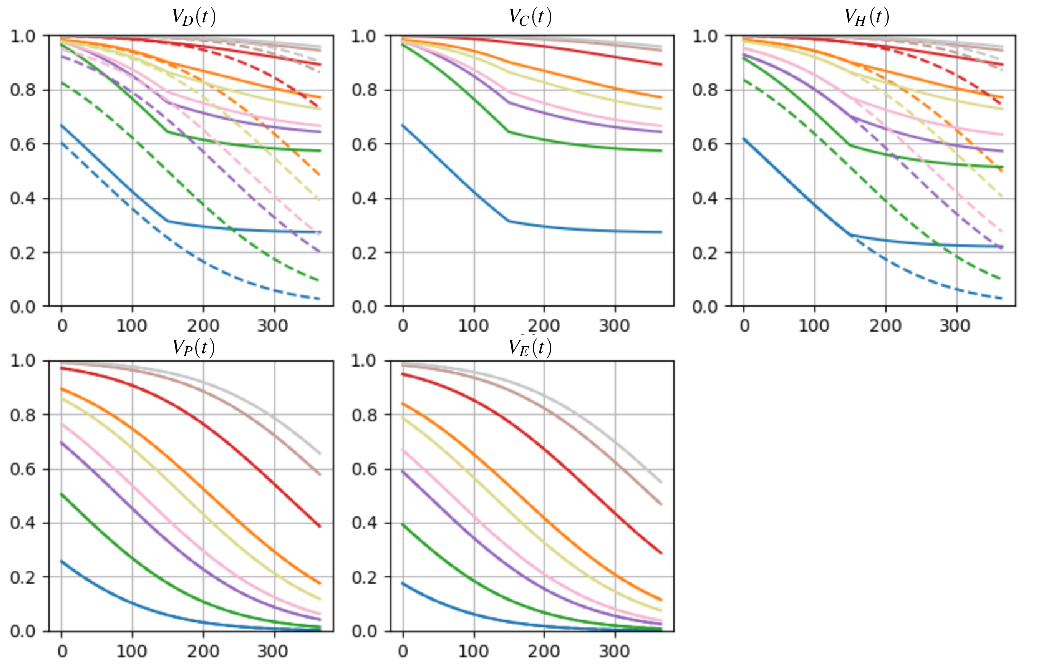}
    \caption{Plot of the target immunity versus time for different disease states, shown with solid curves, compared to the baseline immunity from Golding's model, shown with dashed curves. In increasing order of the protection that they provide: the blue, green, purple, pink, yellow, orange, red, brown, and grey curves give the immunity provided by one dose, two doses, three doses, four doses, previous infection, one dose with previous infection, two doses with previous infection, three doses with previous infection and four doses with previous infection, respectively.}
    \label{fig:target_VE_with_NG}
\end{figure}

Here we show the results of constructing a target immunity dataset. The source code used to implement our assumptions and construct the dataset is publicly available \cite{GraymanicomFitWaningImmunityScripts}. We compare the target immunity to the baseline model in Figure \ref{fig:target_VE_with_NG}. The most significant changes are the inclusion of a critical care state, the reduced long-term waning of immunity against severe disease given infection, and the assumption that the protection against death is at least as large as the protection against critical care.

\subsection{Parameter fitting} \label{subsec:parameterisation_results}

We split our numerical investigation into the relationship between $v$ and $V_E$ into two parts: First, we confirm that our theoretical estimates for the relationship between $v$ and $V_E$ in households of size two, given in (\ref{eq:VE_from_HSAR}), (\ref{eq:HSAR_v}) 
and (\ref{eq:HSAR_u}), are realized in our model. Second, we extend our numerical investigations to households of any size on a synthetic network household representative of Aotearoa New Zealand. In both sets of simulations, we restrict the spread to households only.

For our first set of simulations, we seed $28139$ households of size two with an infected individual, five times, to produce $140695$ independent simulations of pairs of contacts. We vaccinate every susceptible individual with constant protection from exposure per infection attempt $v$ and give everyone full immunity from previous infections so that no one can be reinfected. We set run the simulation until everyone has either recovered or died. This prevents susceptible contacts from escaping infection due to the simulation terminating while their housemate was still infectious. To account for differences in model performance as the underlying infectivity changes, let $\epsilon_X$ be a positive constant dependent on the infectious state $X\in\{A,P,I\}$, $k$ be an integer, $\beta_0$ be a constant representing the minimum infection rate and write $\beta_X= k \epsilon_X\beta_{0}$. For different values of $k$, we simulate the HSAR $H(v)$. We then calculate $V_E$ using (\ref{eq:VE_from_HSAR}). For each pair of parameters, $k$ and $v$, we run five simulations, keeping all other parameters the same. Complete parameter sets are available on request. 

\begin{figure}[th!]
    \centering
    \begin{subfigure}{0.49\linewidth}
        \includegraphics[width=\linewidth]{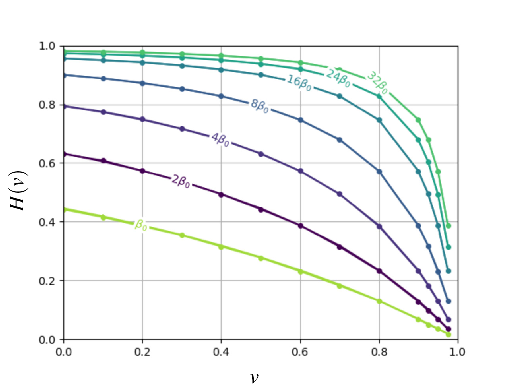}
    \end{subfigure}
    \hfill
    \begin{subfigure}{0.49\linewidth}
        \includegraphics[width=\linewidth]{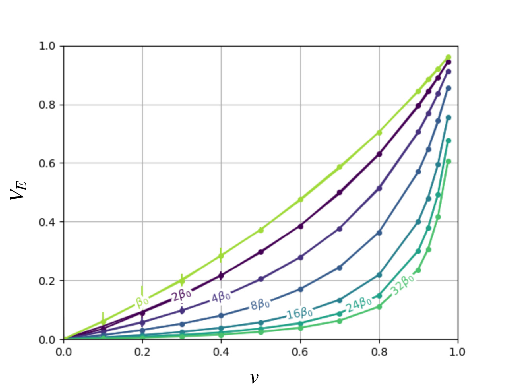}
    \end{subfigure}
    \caption{Plots of the protection against disease acquisition per infection attempt $v$ against the household secondary attack rate (left) and the cumulative protection against acquisition $V_E$ (right) within households of size two. Results from our numerical simulations are shown with circles, while theoretical results are shown with solid curves. The simulation results cover different infection rates $k\beta_0$, where $k\in\{1,2,4,8,16,32\}$, and immunity values $v\in[0,1]$.}
    \label{fig:SAR_and_VE_hsize_2}
\end{figure}

Simulation results are shown with circles in Figure \ref{fig:SAR_and_VE_hsize_2} while theoretical results are shown with solid lines. Note that the probabilities of disease progression $\eta_h$ and $\eta_p$ in (\ref{eq:P_I}), (\ref{eq:HSAR_v}) and (\ref{eq:HSAR_u}) depend on the age and ethnicity of each individual. When calculating the theoretical values we use the population averages. We observe that as $v$ increases the HSAR decreases, while as the infectivity increases the HSAR increases. Furthermore, we see that when the infectivity is small, the relationship between $v$ and $V_E$ is almost linear, since with fewer infection attempts the overall protection will be similar to the protection per infection attempt. For high infectivity, the relationship looks like a power relationship. 

\begin{figure}[t!]
    \centering
    \begin{subfigure}{0.49\linewidth}
        \includegraphics[width=\linewidth]{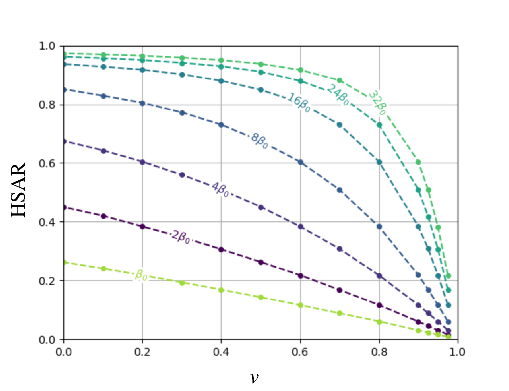}
    \end{subfigure}
    \hfill
    \begin{subfigure}{0.49\linewidth}
        \includegraphics[width=\linewidth]{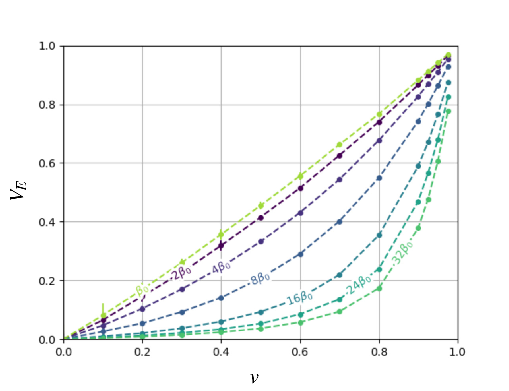}
    \end{subfigure}
    \caption{Plots of the protection against acquisition per infection attempt $v$ vs the household secondary attack rate (left) and the cumulative protection against acquisition $V_E$ (right). Simulation results are shown with circles and are connected by dashed curves. These are taken for different infection rates across all household sizes of the simulation.}
    \label{fig:SAR_and_VE_average_hsize}
\end{figure}

The simulation results match the theoretical values calculated using (\ref{eq:VE_from_HSAR}), which validates us using a numerical approach for larger households, that is, the second experiment. We do this using a sample of $65489$ households randomly drawn from a representative network of New Zealand. The average household size in this sample is $2.65$ and households of size one are excluded. We vary $k$ and $v$ in the same way as before and run simulations for each combination five times. The numerical results are shown in Figure \ref{fig:SAR_and_VE_average_hsize}. 

 In both Figures \ref{fig:SAR_and_VE_hsize_2} and \ref{fig:SAR_and_VE_average_hsize} we show $95\%$ confidence intervals for our simulation results with vertical lines, however these are mostly too small to see on the plots. For households of size two, the number of infected individuals within each simulation has a binomial distribution, while the means of our five simulations are normally distributed, allowing us to calculate the $95\%$ CI with relative ease. However, in general, the number of infected individuals is neither binomially distributed nor can be approximated by a binomial distribution when the HSAR or household sizes are large \cite{sharkerEstimatingInterpretingSecondary2021}. This is because, even if the probability that a susceptible individual gets directly infected by the infectious seed is small, there may be so many chains of transmission that result in that seed case getting infected that their overall probability of infection remains large. However, in our case, less than half of our households have a size greater than two, with a mean of $2.65$. Thus, we use the binomial approximation.

 We use these numerical results to determine which values $v(t)$ produce the target values $V_E(t)$ for a given infectivity $\beta_X$, by simulating disease spread over households for $t$ ranging up to one year. We do this by geometrically interpolating between different values of $k$, fitting curves of the form $
(\ref{eq:HSAR_v})$ to the data and using a root-finding algorithm.  In some instances, more than one $v$ may give the same $V_E$. In that case, we first discard values of $v$ if $v<0$ or $v>1$, then discard those that return $H(v)>H_0$, and then we discard those values of $v$ that change by large amounts when $V_E$ is perturbed slightly. In this way, we construct a set of unique values for $v(t)$ from target values $V_E(t)$ such that we can fit $L_E$, $k_E$ and $T_E$ using equation (\ref{eq:v}). The full household contact network is available online \cite{emilyharveyHouseholdSyntheticNetwork2024}.
 
\begin{figure}[th!]
    \centering
    \includegraphics[width=0.95\linewidth]{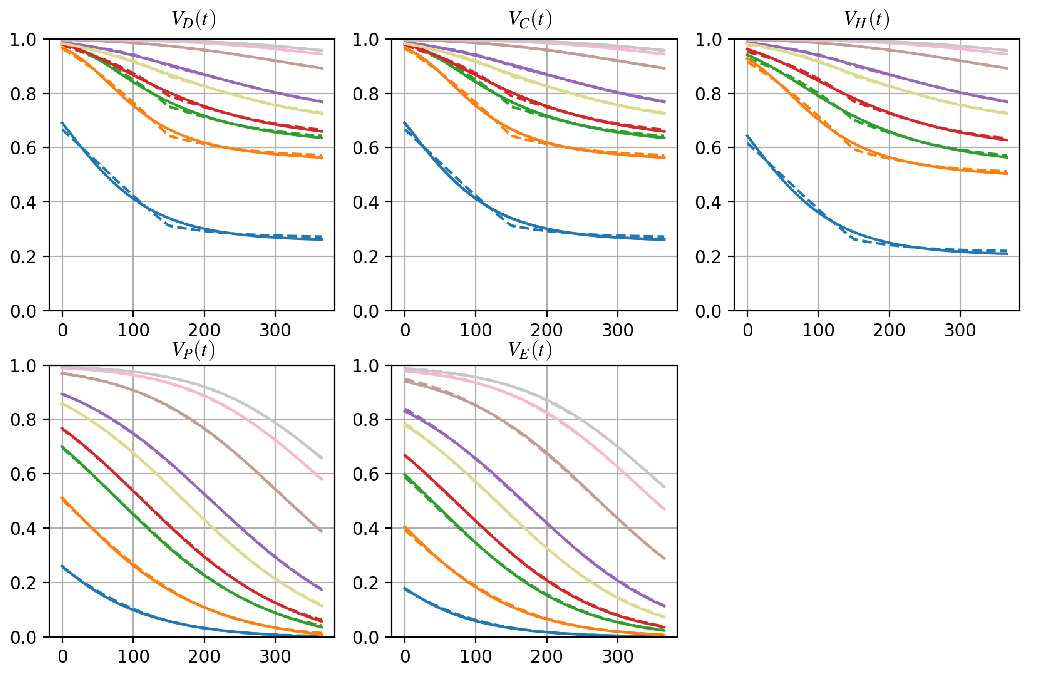}
    \caption{The protection against different disease states observed in our model. The dashed curves show the target immunities that we fit to, shown with solid curves in Figure \ref{fig:target_VE_with_NG}, while the solid curves are the immunity levels that result from our method of parameter-fitting. In order of the protection that they provide: the blue, green, purple, pink, yellow, orange, red, brown, and grey curves give the protection provided by one dose, two doses, three doses, four doses, previous infection, one dose with previous infection, two doses with previous infection, three doses with previous infection and four doses with previous infection, respectively. In cases where the dashed curves are not visible, they overlap with the corresponding solid curves. Note that the solid curves that result from our parameterization are smooth, whereas our target values were non-smooth at $5$ months.}
    \label{fig:resultant_protection}
\end{figure}

With $f_E(t;L_E,k_E,T_E)$ determined we fit our remaining parameters following (\ref{eq:recursive}), and we repeat this process for each of the nine different immunities. The NCM uses this parameterization to calculate the associated transition probabilities between disease states, and we show the resulting immunity levels that will be observed in model results in Figure \ref{fig:resultant_protection}. The immunity values that our model reproduces are close to the values contained in our target dataset. 

\section{Discussion} \label{sec:discussion}

In this paper, we described a method for incorporating waning immunity from leaky vaccines or previous infections into an agent-based model, which uses numerical simulations to take the underlying model assumptions and disease characteristics into account. To do this, we constructed a target dataset of immunity against different disease outcomes from vaccination with or without previous infection and how it wanes over time. We fitted logistic curves to the target values to give a parsimonious parameterization of waning immunity. Since the probability of a susceptible individual acquiring the disease depends on the force of infection, their contact structure, and other modelling assumptions, we used numerical simulations of the HSAR to determine the relationship between the reduction in the probability of infection from each infection attempt an infectious individual makes on a susceptible individual and the overall protection against disease acquisition across many contacts for different values of the force of infection. We confirmed our simulations with theoretical calculations of a simplified scenario where households were all of size two. While the approach is useful for all infectious diseases with leaky vaccines, we demonstrated the approach with a model of the BA.1 and BA.2 strains of the Omicron variant which was dominant in New Zealand during 2022 when the population had little to no immunity from exposure to previous variants \cite{jelleyTracingHouseholdTransmission2024}. 

This approach is useful for three reasons. Firstly, it is a relatively straightforward way of considering the model's assumptions, which is critical for choosing the model parameters that result in realistic values of immunity. Secondly, it better represents the reality of leaky vaccines, since observed vaccine effectiveness depends on the force of infection and the resulting number of infection attempts. For example, vaccines appeared to be less effective against more infectious COVID-19 variants, but this may have resulted from the increased force of infection while the underlying protection provided against each infection attempt remained the same. Thirdly, it makes the model more applicable to a range of real-world settings since biological or environmental changes that impact the force of infection (changes in $\beta_X$) and the observed vaccine effectiveness (changes in $V_E(t)$) can easily be accommodated for by using existing simulation results to choose new parameters describing the protection against acquisition per infection attempt $v(t)$, and, from there, the remaining immunity parameters.

This work has a few limitations. Firstly, our simulations do not fully account for variations in individual attributes. In reality, the proportions of individuals who progress down different disease pathways depends on their age and ethnicity. Thus, different values of vaccine effectiveness could be observed even when those different groups have the same protection from each infection attempt. Additionally, observations suggest that the protection against acquisition and severe disease also vary with age \cite{priddyImmunogenicityBNT162b2COVID192022,baumHighVaccineEffectiveness2022}. These two factors could be combined create separate immunity parameters for different age groups. However, for simplicity, and due to the limited amount of data across different age groups for the dominant variants at the time, we did not do this. Secondly, in these simulations we ignored the effect of non-pharmeceutical interventions such as testing, tracing and isolating, which would affect real-world measurements of VE in non-household settings. 

A final limitation of our model is that the HSAR and resulting measurement of VE in households of different sizes does not represent reality. Since our model assumes a frequency-dependent transmission, each infectious individual will have the make the same number of infection attempts with other individuals in their household regardless of how many individuals share that household with them. (This was confirmed in other simulations and is suggested by Figures \ref{fig:SAR_and_VE_hsize_2} and \ref{fig:SAR_and_VE_average_hsize}, which show that HSAR decreased as the average household size of the simulation increased). In reality, the HSAR increases as the size of the household increases, since there are more contacts and possible transmission chains. This may have equity implications since it is known that in New Zealand and elsewhere different ethnic groups have different average household sizes \cite{wingAssociationHouseholdComposition}. However, this limitation of our model is a strength of the methodology, since we reproduce realistic target immunities despite model assumptions that, in this context, are unrealistic, and emphasises the usefulness of our approach. 

\subsection*{Data availability statement}

The network and code used to run the simulations are available at \cite{cobin_documentation} and \cite{emilyharveyHouseholdSyntheticNetwork2024}, respectively. The code and data used to contruct the target immunity values and fit parameters to them are available at \cite{figshare_data} and \cite{GraymanicomFitWaningImmunityScripts} respectively. 

\section{Acknowledgements}
We acknowledge the contributions of Frankie Patten-Elliott, Steven Turnbull, James Gilmour, and the broader Covid-19 Modelling Aotearoa team for their work on the constructing the individual contact network \cite{emilyharveyHouseholdSyntheticNetwork2024} and the network contagion model \cite{cobin_documentation}. We thank Janine Paynter for fruitful discussions regarding the robustness of immunity against severe disease. This work was supported by grants from the New Zealand Ministry of Health (Manatū Hauora), the NZ Ministry for Business, Innovation, and Employment (MBIE), and the NZ Health Research Council (HRC).

\section{Declaration of interest}
Competing interests: The authors declare none.

\clearpage
\printbibliography

\end{document}